\def\papertitle{Sound Matching with a Differentiable Karplus-Strong Algorithm}
\def\paperauthorA{Pablo Tablas de Paula}
\def\paperauthorB{David Marttila}
\def\paperauthorC{Rodrigo Díaz}
\def\paperauthorD{Irán Román}
\def\paperauthorE{Emmanouil Benetos}
\def\paperauthorF{Joshua D. Reiss}

\documentclass[twoside,a4paper]{article}
\usepackage{etoolbox}
\usepackage[]{dafx26v3}
\usepackage{amsmath,amssymb,amsfonts,amsthm}
\usepackage{siunitx}
\usepackage{euscript}
\usepackage[T1]{fontenc}
\usepackage[utf8]{inputenc}
\usepackage{ifpdf}
\usepackage[english]{babel}
\usepackage{caption}
\usepackage{subfig} 
\usepackage{color}
\usepackage{booktabs}
\usepackage{lipsum}

\usepackage{booktabs}
\usepackage{enumitem} 

\usepackage{glossaries}
\usepackage{acronym}
\newacronym{ksa}{KSA}{Karplus-Strong Algorithm}
\newacronym{nn}{NN}{Neural Network}
\newacronym{ddsp}{DDSP}{Differentiable Digital Signal Processing}
\newacronym{dwg}{DWG}{Digital Waveguide}
\newacronym{p_loss}{$\mathcal{P}_{loss}$}{Parameter Loss}
\newacronym{mss}{$\mathcal{L}_{MSS}$}{Multi-Scale Spectral Loss}
\newacronym{f_0}{$f_0$}{Fundamental Frequency}
\newacronym{sot}{$\mathcal{L}_{SOT}$}{Spectral Optimal Transport}
\newacronym{iir}{IIR}{Infinite Impulse Response}
\newacronym{fir}{FIR}{Finite Impulse Response}

\newacronym{stft}{STFT}{Short-Time Fourier Transform}
\newacronym{mfcc}{MFCC}{Mel-Frequency Cepstral Coefficients}
\newacronym{dtw}{DTW}{Dynamic Time Warping}
\newacronym{rms}{RMS}{Root Mean Square}
\newacronym{mmd}{MMD}{Maximum Mean Discrepancy}
\newacronym{kad}{KAD}{Kernel Audio Distance}
\newacronym{rbf}{RBF}{Radial Basis Function}
\newacronym{hn}{HpN}{Harmonics plus Noise}
\newacronym{lti}{LTI}{Linear Time Invariant}
\newacronym{ltv}{LTV}{Linear Time Variant}

\newacronym{tcn}{TCN}{Temporal Convolutional Network}

\newacronym{ola}{OLA}{Overlap-add}
\newacronym{ste}{STE}{Straight Through Estimator}

\input glyphtounicode
\ninept

\newcounter{numauth}
\newcounter{listcnt}
\newcommand\authcnt[1]{\ifdefined#1 \stepcounter{numauth} \fi}

\newcommand\addauth[1]{
\ifdefined#1 
\stepcounter{listcnt}
\ifnum \value{listcnt}<\value{numauth}
\appto\authorslist{, #1}
\else
\appto\authorslist{~and~#1}
\fi
\fi}
\authcnt{\paperauthorB}
\authcnt{\paperauthorC}
\authcnt{\paperauthorD}
\authcnt{\paperauthorE}
\authcnt{\paperauthorF}
\authcnt{\paperauthorG}
\authcnt{\paperauthorH}
\authcnt{\paperauthorI}
\authcnt{\paperauthorJ}
\def\authorslist{\paperauthorA}
\addauth{\paperauthorB}
\addauth{\paperauthorC}
\addauth{\paperauthorD}
\addauth{\paperauthorE}
\addauth{\paperauthorF}
\addauth{\paperauthorG}
\addauth{\paperauthorH}
\addauth{\paperauthorI}
\addauth{\paperauthorJ}

\usepackage{times}

\newif\ifpdf
\ifx\pdfoutput\relax
\else
   \ifcase\pdfoutput
      \pdffalse
   \else
      \pdftrue
   \fi
\fi

\ifpdf 
  \usepackage[pdftex,
    pdftitle={\papertitle},
    pdfauthor={\authorslist},
    pdfsubject={Proceedings of the 29th International Conference on Digital Audio Effects (DAFx26)},
    colorlinks=false, 
    bookmarksnumbered, 
    pdfstartview=XYZ 
  ]{hyperref}
  \usepackage[pdftex]{graphicx}
\else 
  \usepackage[dvips]{epsfig,graphicx}
  \usepackage[dvips,
    pdftitle={\papertitle},
    pdfauthor={\authorslist},
    pdfsubject={Proceedings of the 29th International Conference on Digital Audio Effects (DAFx26)},
    colorlinks=false, 
    bookmarksnumbered, 
    pdfstartview=XYZ 
  ]{hyperref}
\fi
\usepackage[hypcap=true]{caption}
\title{\papertitle}

 \affiliation
 {\paperauthorA\sthanks{Supported by the UKRI Centre for Doctoral Training in Artificial Intelligence and Music (grant EP/S022694/1.)\vspace{3mm}}\/, \paperauthorB\/, \paperauthorC\-, \\ \paperauthorD\/, \paperauthorE\/ and \paperauthorF\/}
 {\href{https://www.c4dm.eecs.qmul.ac.uk/}{Centre For Digital Music} \\ Queen Mary University Of London \\ London, UK\\
 {\tt \href{mailto:p.tablasdepaula@qmul.ac.uk}{p.tablasdepaula@qmul.ac.uk}}
 }
\begin{document}
\sloppy
\ifpdf 
  \DeclareGraphicsExtensions{.png,.jpg,.pdf}
\else  
  \DeclareGraphicsExtensions{.eps}
\fi


\maketitle

\begin{abstract}
We present a self-supervised, event-based sound matching model using a differentiable extended Karplus-Strong algorithm. To avoid relying on external onset and fundamental frequency detectors, we explore training methodologies combining parameter losses on synthetic data with audio losses. We demonstrate that time-domain fractional delay interpolation provides gradient accuracy comparable to frequency-sampling while avoiding time-aliasing in highly resonant time-varying scenarios. Through systematic gradient analysis, we reveal that standard spectral losses provide no meaningful directional gradients for onset times, heavily degrading joint training. Training exclusively with parameter losses on synthetic data effectively learns fundamental frequency, timbral parameters, and onset times, but struggles to generalise to monophonic studio recordings of plucked guitar. External detectors combined with audio losses generalise best, isolating the model to timbre optimisation. While our Karplus-Strong decoder recovers interpretable parameters and naturally captures the transient characteristics of plucked guitar, Harmonics plus Noise baselines yield higher reconstruction fidelity by most metrics.
\end{abstract}

\glsresetall

\section{Introduction}
\label{sec:intro}
The \gls{ksa} offers an efficient method to simulate plucked instruments \cite{Karplus:1983:Digital}. Originally conceived as a self-changing wavetable, Jaffe and Smith re-interpreted it as a fed-back delay line with additional filtering to simulate dispersion \cite{Jaffe:1983:Extensions}, an insight that would inspire the wider field of \gls{dwg} Synthesis and enable physical modelling in real-time \cite{Tablas:2026:Four}.

Engel et al.\ demonstrated that DSP parameters can be optimised end-to-end with neural networks via \gls{ddsp} \cite{Engel:2020:DDSP}, a self-supervised analysis-by-synthesis paradigm in which controls are learned directly from audio, sparking growing interest in extending the technique beyond the original \gls{hn} decoder \cite{Caspe:2022:DDX7, Sudholt:2023:Vocal}. The \gls{ksa} is a natural candidate: it requires far fewer parameters than \gls{hn} and naturally captures transient characteristics. In the online supplement to Hayes et al., audio losses were used to optimise the decay of a frequency-sampling \gls{lti} implementation of \gls{ksa} \cite{Hayes:2024:Review}. More recently, we extended this to a fully time-domain implementation in which decay, timbre, pluck-position, and dynamics are free to vary at every frame \cite{Tablas:2025:Differentiable}. 

However, both studies relied on external $f_0$ and onset detectors; the first are prone to octave errors due to pitch fluctuations around transients, while the latter introduce time-shifts, missed onsets and confounds string-fret interactions with plucks. Additionally, this frame-based parameter prediction is unnecessary for discrete plucks: assuming expressive techniques such as vibrato or pitch bends are absent, a pluck's damping, decay, and $f_0$ are constant over its note.

Optimising these parameters by gradient descent requires fully differentiable architectures. In the frequency domain, fractional delays become continuous phase shifts \cite{Pei:2012:Closed, Mezza:2025:Differentiable, Carson:2026:Gradient}, but time-aliasing is severe when impulse responses exceed the FFT frame size \cite{Dal:2025:Flamo}. Time-domain interpolation avoids this and is now computationally viable through parallelised automatic differentiation \cite{Yu:2025:Accelerating}; however, its gradient behaviour under spectral losses is unexplored. 

Such losses are notoriously unreliable for pitch \cite{Turian:2020:Sorry} and, as we show, give no useful gradients for onset times. Recent \gls{ddsp} work stabilises frequency prediction with \gls{p_loss}, the distance between parameters of synthetic audio made by the synthesiser in question \cite{Engel:2020:Self, Masuda:2023:Improving}; we extend this to onset supervision. Our broader aim is a self-supervised event-based model: one that recovers the discrete structure of a performance (how many plucks, when, and at what pitch) without hand-annotated onsets or $f_0$, jointly with the string timbre, deriving its supervision instead from the differentiable synthesiser and freely-labelled synthetic data.

Our contributions are threefold:
\begin{itemize}[topsep=2pt,itemsep=2pt,parsep=0pt,partopsep=0pt,leftmargin=*]
    \item We show that delay line length is differentiable in the time domain via Lagrange Interpolation, with gradient behaviour comparable to frequency sampling and without time-aliasing artefacts.
    \item We perform a systematic gradient analysis showing that standard spectral losses fail to provide meaningful directional gradients for onset times, heavily degrading joint audio-parameter optimisation.
    \item We demonstrate that \gls{p_loss} on synthetic data effectively learns $f_0$, timbre, and onset times, while external detectors combined with audio losses generalise best to real-world data.
\end{itemize}

\section{Background}
\label{sec:background}
\subsection{Karplus-Strong Synthesis}\label{sec:background:karplus}

\begin{figure}[t]
    \centering
    \includegraphics[width=\columnwidth]{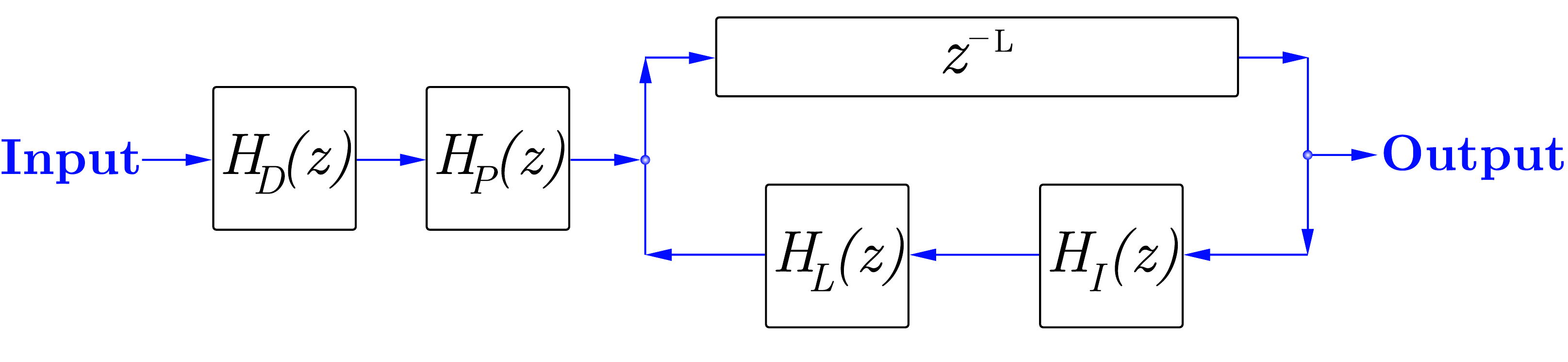}
    \caption{An extended Karplus-Strong Algorithm.}
    \label{fig:e_ksa}
\end{figure}

The \gls{ksa} circulates a short broadband noise burst, typically one period of the desired \gls{f_0}, through a digital delay line in a feedback loop, the burst and delay line being analogous to pluck and string. The loop delay $L$ (in possibly fractional samples) is the round-trip time of a transversal wave along the string, setting the perceived \gls{f_0} via
\begin{equation}
L = \frac{f_s}{f_0},
\end{equation}
where $f_s$ is the sampling frequency; this assumes delay-free filters, with the loop filter's phase delay later compensated in $D$. In the feedback path, a \emph{loop filter} $H_L(z)$ models frequency-dependent string losses. The original wavetable used a two-point average $H_L(z)=\frac{1}{2}(1 + z^{-1})$; we adopt a more modern approach proposed by Välimäki et al., which favours the expressiveness of \gls{iir} filters:
\begin{equation}
\label{eq:loop_filter}
    H_L(z) = g\frac{1 + a_1}{1 + a_1z^{-1}}.
\end{equation}
Here $H_L(z)$ is implemented as a first-order all-pole filter and re-parameterised to provide separate, decoupled control over the overall decay time via the DC gain $g$, and frequency-dependent damping via the cutoff coefficient $a_1$ \cite{Valimaki:1996:Physical}.

Because the ideal delay length $L$ rarely resolves to an integer number of samples, an interpolation filter $H_I(z)$ in the feedback path makes the delay fractional. All-pass filters avoid the strong low-pass effect of linear interpolation \cite{Jaffe:1983:Extensions}, but their recursive nature makes them prone to transients when the delay changes. We instead use \emph{Lagrange Interpolation}, whose fractional delay is far more uniform across frequency, with a maximally flat magnitude response in the lower spectrum \cite{Laakso:1996:Splitting}. The coefficients $h[n]$ for the Lagrange interpolator are computed as:
\begin{equation}
h[n] = \prod_{\substack{k=0 \\ k \neq n}}^{N} \frac{D - k}{n - k}, \quad n=0, 1, \dots, N,
\end{equation}
where $N$ is the order of the \gls{fir} filter and $D$ the desired delay. We use $N=5$, which keeps the fractional delay accurate while limiting passband distortion \cite{Laakso:1996:Splitting}. The final Lagrange interpolator has the form:
\begin{equation}
\label{eq:lagrange_fir}
H_I(z) = \sum_{n=0}^{N} h[n] z^{-n}.
\end{equation}

Since $H_L(z)$ adds its own frequency-dependent phase delay, the fractional delay $D$ fed to the Lagrange interpolator must absorb both the non-integer remainder of $L$ and the loop-filter phase delay $P_L(f_0)$ \cite{Jaffe:1983:Extensions}:
\begin{equation}
    D = L - \lfloor L \rfloor - P_L(f_0), \quad \text{where} \quad P_L(f_0) = -\frac{\angle H_L\left(e^{j\omega_0}\right)}{\omega_0}.
\end{equation}
Before entering the delay line, the noise burst is shaped by two excitation filters \cite{Jaffe:1983:Extensions}. The \emph{pluck position filter} models where along the string it is struck: plucking at a fractional length imposes nodes that suppress certain harmonics, captured by a feedforward comb $H_P(z) = 1 - z^{-p}$ with $p$ proportional to the pluck position. Being feedforward and applied once, its fractional delay reduces to linear interpolation:
\begin{equation}
\label{eq:pluck}
H_P(z) = 1 - z^{-p_i}\!\left((1 - p_f) + p_f\, z^{-1}\right),
\end{equation}
where $p_i = \lfloor p \rfloor$ and $p_f = p - \lfloor p \rfloor$.

Lastly, since stronger impacts introduce more high-frequency energy, Jaffe and Smith model intensity with a \emph{dynamics filter} \cite{Jaffe:1983:Extensions}, which acts perceptually as a strike intensity rather than a distance control:
\begin{equation}
H_D(z) = \frac{1 - R}{1 - Rz^{-1}},
\end{equation}
where $R\in[0,1]$ is computed from the fundamental frequency $f_0$ and the desired dynamic level to keep perceived loudness uniform across the range (derivation in \cite{Jaffe:1983:Extensions}). The complete extended algorithm is shown in Figure \ref{fig:e_ksa}.

\subsection{Synthesiser Sound Matching}
Sound matching is the inverse of synthesis: given a target sound, recover the parameters that reproduce it. For physical models like the \gls{ksa} this is also analytical; the predicted parameters give interpretable insight into the object's physical properties \cite{Riionheimo:2003:Parameter}. Early, costly evolutionary methods \cite{Riionheimo:2003:Parameter} gave way to supervised \gls{nn} regression \cite{Gabrielli:2017:Introducing} minimising \gls{p_loss}; however, \gls{p_loss} optimises numerical rather than acoustic accuracy, often yielding inconsistent perceptual reconstructions \cite{Masuda:2023:Improving}.

\subsection{Differentiable Digital Signal Processing}
\gls{ddsp} implements signal processors as differentiable operations, turning the synthesiser into an interpretable decoder so an audio-reconstruction loss can backpropagate to its controls and train end-to-end \cite{Engel:2020:DDSP}. Its standard objective, \gls{mss}, measures $L_1$ distance between amplitude spectrograms, effective for timbre but unreliable for pitch \cite{Turian:2020:Sorry}, often needing external \gls{f_0} predictors \cite{Engel:2020:DDSP}. Recent work mitigates this via Wasserstein-based \gls{sot} for frequency shifts \cite{Torres:2024:Unsupervised}, revised \gls{mss} configurations \cite{Schwar:2023:Multi}, and \gls{p_loss} pre-training \cite{Masuda:2023:Improving}.

\subsection{Gradient-Based Optimisation for Recursive Filters}\label{sec:background:grad}
Naive time-domain implementation of recursive filters requires sample-by-sample calculation, generating prohibitively large computational graphs. Two complementary approaches restore parallelism.

\begin{figure*}[t!]
  \centering
  \includegraphics[width=\textwidth, height=0.25\textheight, keepaspectratio]{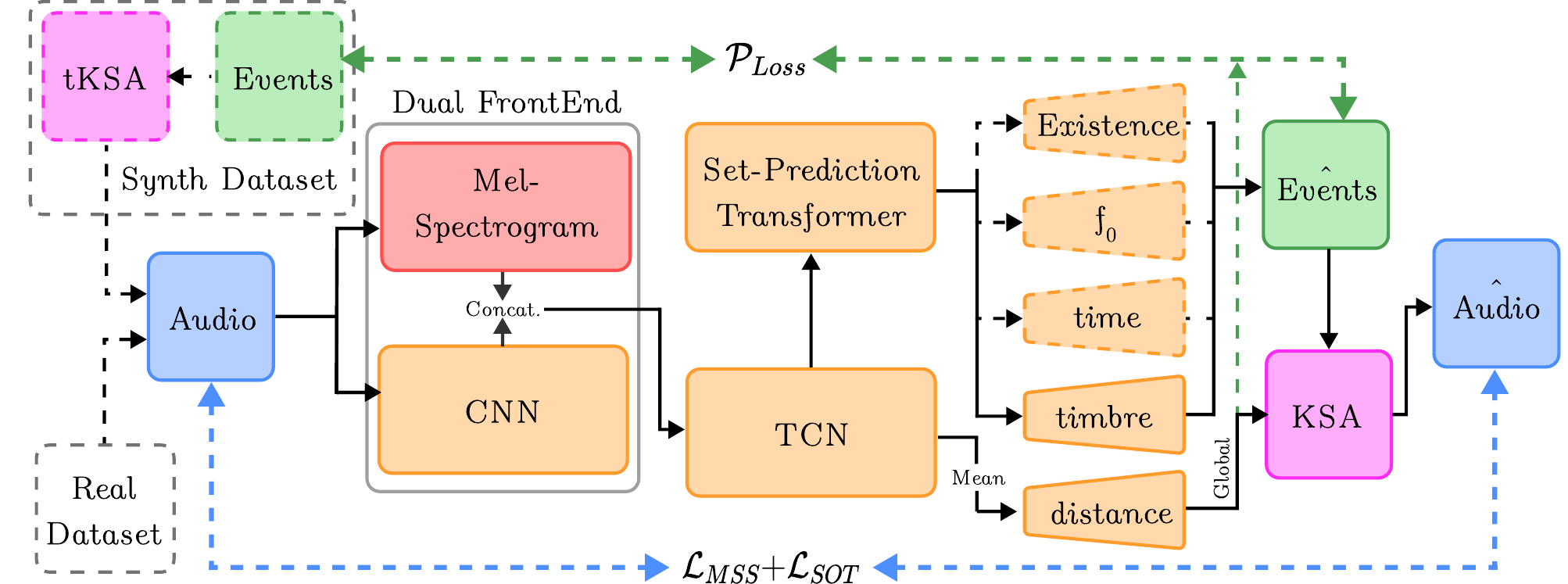}
  \caption{Proposed sound matching pipeline. Dashed lines indicate optional paths defining our three training regimes: $\mathcal{P}$-Only (end-to-end prediction trained with \gls{p_loss} uniquely on Synth Data), $\mathcal{P}$+Audio (end-to-end prediction trained with both \gls{p_loss} and audio losses), and Audio-Only (trained solely with audio losses, bypassing the existence, $f_0$ and onset time parameter heads with external detectors).}
  \label{fig:model}
\end{figure*}

The \emph{frequency sampling method} evaluates the filter response over $M$ linearly-spaced frequencies:
\begin{equation}
z_M = \left[ e^{j\pi\frac{0}{M-1}}, e^{j\pi\frac{1}{M-1}}, \dots, e^{j\pi} \right]
\end{equation}
yielding a parallelisable \gls{fir} approximation via DFT ratio of feedforward and feedback coefficients. Arbitrarily complex filter structures (series connections and recursive loops) are composed through elementwise products and matrix inversions over $z_M$ \cite{Dal:2025:Flamo}, and fractional delays $D$ (in samples) reduce to a smooth and differentiable closed-form phase shift~\cite{Pei:2012:Closed}:
\begin{equation}
\label{eq:freq_delay}
    z_M^{-D} = e^{-j 2\pi k M^{-1} D}.
\end{equation}
A Hermitian symmetry constraint enforces a real-valued impulse response of length $N_\text{FFT} = 2(M{-}1)$; if the true impulse response exceeds this, circular convolution introduces time-aliasing. For time-varying parameters, the \gls{ddsp} literature combines frequency sampling with frame-based \gls{ola}  \cite{Engel:2020:DDSP, Sudholt:2023:Vocal, Carson:2026:Gradient}.

Alternatively, Yu and Fazekas restore native time-domain parallelism by computing exact analytical gradients for recursive filters via the adjoint method, eliminating both memory bottlenecks and the truncation artefacts inherent to frequency sampling \cite{Yu:2025:Accelerating}. Fractional delays in this setting have been handled by backpropagating through discrete interpolators \cite{Churchwell:2025:Combolutional}, but to the knowledge of the authors, not through audio losses.

\subsection{Set-Based Prediction}
We base our event encoder on the Sound Event Detection Transformer (SEDT)~\cite{Ye:2021:Sound}, an event-based, end-to-end model that frames detection as direct set prediction: a fixed set of learnable queries is matched to the ground-truth events in a single pass, regressing each event's boundaries without frame-wise post-processing or autoregressive decoding. We adapt this paradigm to jointly estimate discrete pluck events and their physical parameters.

\section{Differentiable Karplus-Strong}
\label{sec:method}
Because the \gls{ksa} is a recursive, energy-dissipating system, predicting parameters on a continuous frame-by-frame basis can cause errors to accumulate through the feedback loop. Provided no extended techniques are present in the data, it is more robust to use an event-based model, representing acoustic variations as discrete note-level events with constant damping, decay, and $f_0$. Figure \ref{fig:model} illustrates the complete pipeline, consisting of an Event-Based Encoder (§\ref{subsec:event}) and two differentiable \gls{ksa} decoder variants (§\ref{subsec:ksa_dec}): a Time-Domain (tKSA) and a Frequency-Sampling (fKSA) decoder.

\subsection{Event-Based Encoder}\label{subsec:event}
The encoder maps a monophonic waveform to a fixed-size set of $N{=}10$ event parameter slots through four stages. We set $N{=}10$ as headroom: most samples contain only one or two plucks, but a few NSynth recordings (e.g.\ tempo-synced ones; \cite{Engel:2017:Neural}) contain more.

\textbf{Dual-Frontend Feature Extractor}. Inspired by \cite{Shier:2023:Differentiable}, the audio (4\,s, $f_s{=}16$\,kHz; §\ref{sec:data}) is processed in parallel by a log-magnitude Mel filterbank (1024-point FFT, 256-sample hop, 128 bands, 32--8000\,Hz) for $f_0$ and harmonic structure and a learnable 1D CNN (4 layers, up to 64 channels, stride 4) for phase-sensitive transients; each is followed by batch norm and ReLU, then truncated to a common time axis and concatenated.

\textbf{Temporal Convolutional Network}. The concatenated features are projected to $d_\text{model}=64$ and processed by a 7-block \gls{tcn} with exponentially increasing dilations $(1, 2, \dots, 64)$, building a global contextual memory representation.

\textbf{Set-Prediction Transformer}. $N=10$ learnable query embeddings attend to the \gls{tcn} memory via 3 standard Transformer decoder layers (8-head self- and cross-attention, GELU feed-forward), with sinusoidal positional encoding, layer normalisation, and dropout ($p=0.1$).

\textbf{Parameter Projection Heads}. Each query embedding is decoded by shared linear heads predicting: event existence (sigmoid logit); onset time (sigmoid-mapped scalar); \gls{f_0} (soft-argmax over 128 log-spaced bins, 32--2000 Hz); and four timbral parameters: decay ($g$), damping ($a_1$), pluck position ($p$), and dynamic level, each sigmoid-mapped to physical bounds. A separate head predicts a global distance gain from the mean \gls{tcn} representation.

\subsection{\gls{ksa} Decoders}\label{subsec:ksa_dec}

At inference, and for both \gls{p_loss} training and synthetic dataset generation, we use a buffer-based, sample-by-sample, non-differentiable implementation of the \gls{ksa}. Training under audio losses uses a differentiable onset by placing a noise burst via a \gls{ste}: the forward pass snaps to the nearest integer sample, while the backward pass routes gradients through the continuous phase rotation of Eq.~\ref{eq:freq_delay}, giving smooth directional gradients for onset time. This excitation method is paired with the following differentiable KSA implementations:

\textbf{Time-Domain (tKSA)}. Using parallelised backpropagation through all-pole filters \cite{Yu:2024:Differentiable}, the loop filter $H_L(z)$, Lagrange interpolator $H_I(z)$, and integer delay $z^{-L_{int}}$ are combined into a single closed-loop transfer function:
\begin{equation}
\label{eq:tKSA}
    H_{tKSA}(z) = \frac{H_D(z)H_P(z)}{1 - H_L(z)H_I(z)z^{-L_{int}}},
\end{equation}
where $L_{int} = \lfloor L \rfloor$.

\textbf{Frequency Sampling (fKSA)}. Discrete interpolators and integer delays are replaced by the continuous fractional phase shift of Eq.~\ref{eq:freq_delay}:
\begin{equation}
    H_{fKSA}(z_M) = \frac{H_D(z_M)H_P(z_M)}{1 - H_L(z_M)z^{-L}_{M}}.
\end{equation}
Owing to the highly resonant nature of \gls{ksa} (Figure \ref{fig:rt60}), no single $N_{FFT}$ avoids time-aliasing for all configurations while preserving time-varying dynamics. We find $N_{FFT}=2^{14}$ with a 256-sample Hann-windowed hop an effective compromise for the decay and onset timings of our dataset (predominantly a sustained note at 0s with a palm mute at 3s; see appendix \ref{sec:appendix} for sound examples), though time-aliasing still occurs (Fig. \ref{fig:nsynth_tKSA_vs_fKSA}).

\section{Training}
\label{sec:training}
\subsection{Losses}\label{sec:losses}

The total training objective combines two audio losses with a parameter loss, each defined in the subsections below:
\begin{equation}\label{eq:total-loss}
    \mathcal{L} = w_{\text{MSS}}\,\mathcal{L}_{\text{MSS}}
               + w_{\text{SOT}}\,\mathcal{L}_{\text{SOT}}
               + w_{\mathcal{P}}\,\mathcal{P}_{\text{Loss}}.
\end{equation}

\subsubsection{Audio Losses}
We adopt the \emph{Revisited} \gls{mss} of Schw\"{a}r and M\"{u}ller~\cite{Schwar:2023:Multi}, a \emph{vertical} (point-wise) loss~\cite{Torres:2024:Unsupervised} computing the $\ell_2$-distance between log-compressed magnitude spectrograms at six scales (prime window lengths $\{67, 127, 257, 509, 1021, 2053\}$, flat-top window, no zero-padding, $\gamma{=}1$):
\begin{equation}\label{eq:mss}
    \mathcal{L}_{\text{MSS}}
    = \sum_{s=1}^{6}
      \bigl\lVert
        \log(1{+}\gamma\lvert S_s\rvert)
        - \log(1{+}\gamma\lvert \hat{S}_s\rvert)
      \bigr\rVert_2.
\end{equation}
\gls{sot} complements this with a \emph{horizontal} comparison~\cite{Torres:2024:Unsupervised}: within each \gls{stft} frame it measures the squared $2$-Wasserstein cost of transporting magnitude energy \emph{along} the frequency axis, providing directional gradients for frequency shifts where the vertical \gls{mss} fails~\cite{Turian:2020:Sorry}:
\begin{equation}\label{eq:sot}
\begin{aligned}
    \mathcal{L}_{\text{SOT}} &= \sum_{t} W_2^2\!\left(\tilde{S}_t,\,\tilde{\hat{S}}_t\right), \\
    W_2^2(p,q) &= \int_0^1 \bigl(F_p^{-1}(u) - F_q^{-1}(u)\bigr)^2 \, du,
\end{aligned}
\end{equation}
where $\tilde{S}_t$ is the magnitude spectrum of frame $t$ normalised into a probability distribution over frequency and $F^{-1}$ is its quantile (inverse-CDF) function. We adopt the best-performing configuration and weights of Torres et al.\ ($w_{\text{MSS}}{=}0.05$, $w_{\text{SOT}}{=}1.0$)~\cite{Torres:2024:Unsupervised}, but pair them with the smooth Revisited \gls{mss} above.

\subsubsection{\acrfull{p_loss}}\label{sec:event-set-loss}
When ground-truth parameters are available (synthetic data), we optimise a set-prediction loss, following SEDT~\cite{Ye:2021:Sound}. At each step, Hungarian matching assigns the $N$ predicted slots to $M \leq N$ ground-truth events using a cost matrix combining temporal $L_1$ and existence probability; unmatched slots receive zero-padded no-event targets. The matched loss is:
\begin{multline}\label{eq:param-loss}
    \mathcal{P}_{\text{Loss}} = w_e\,\mathcal{L}_e + w_t\,\mathcal{L}_t + w_{f_0}\,\mathcal{L}_{f_0} + w_g\,\mathcal{L}_g \\
    + w_{\text{rt60}}\,\mathcal{L}_{\text{rt60}} + w_p\,\mathcal{L}_p + w_d\,\mathcal{L}_d
\end{multline}
where the weights $(w_e, w_t, w{f_0}, w_g, w_{\text{rt60}}, w_p, w_d)$ are set to $(100, 10, 2, 1, 1, 0.5, 0.5)$, respectively. Existence ($\mathcal{L}_e$) uses sigmoid focal loss~\cite{Lin:2017:Focal} with standard hyperparameters ($\alpha{=}0.5$, $\gamma{=}2$) to handle class imbalance. Onset time, pluck position, dynamic level, and global distance gain ($\mathcal{L}_t, \mathcal{L}_p, \mathcal{L}_d, \mathcal{L}_g$) use $L_1$ losses on sigmoid-activated predictions. Fundamental frequency ($\mathcal{L}_{f_0}$) applies negative log-likelihood over $K{=}128$ log-spaced bins.

\begin{figure}[h!]
    \centering
    \includegraphics[width=\columnwidth]{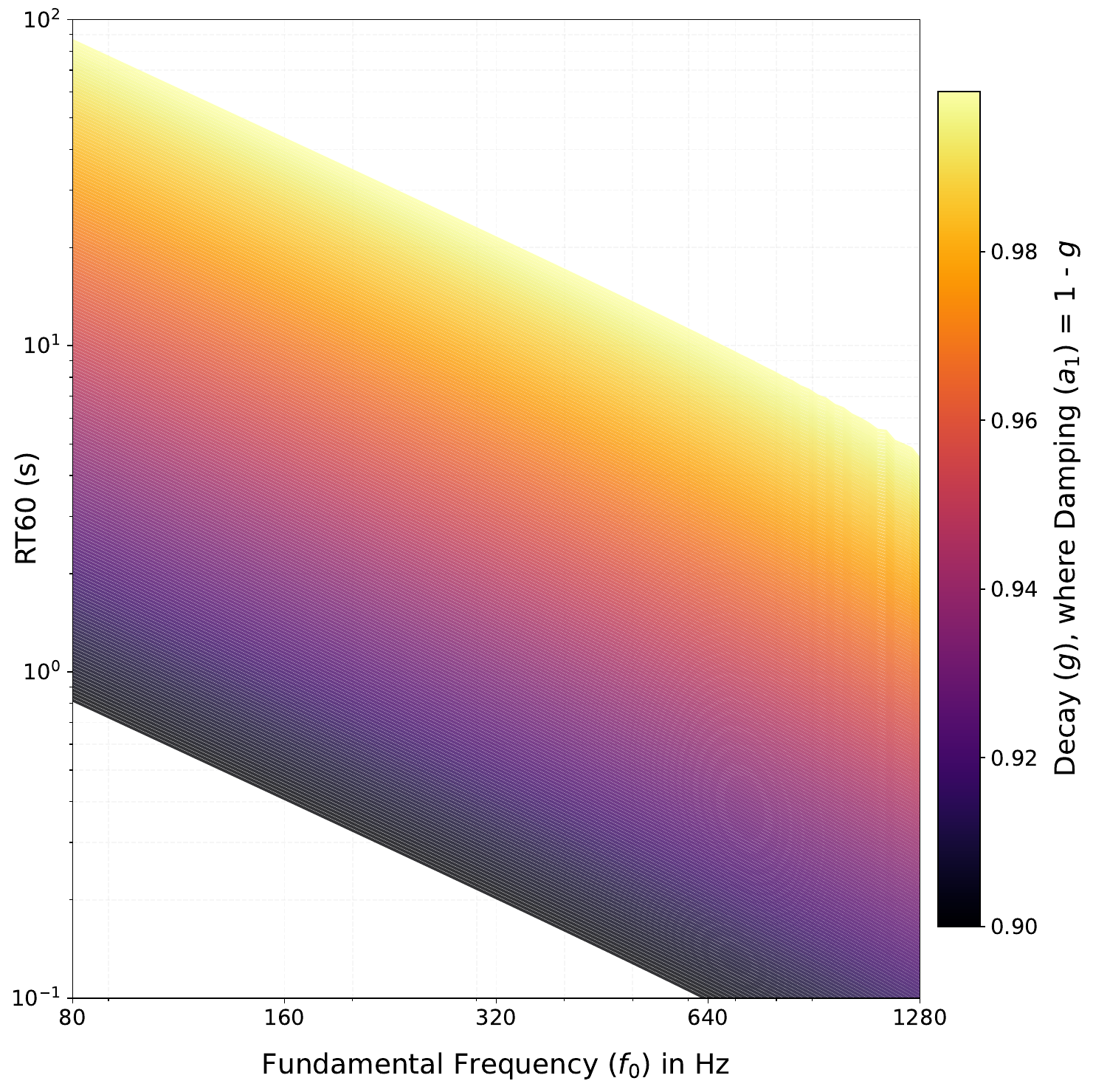}
    \caption{RT60 values across $f_0$, damping ($a_1$), and decay ($g$) in tKSA (without $H_P(z)$ and $H_D(z)$), illustrating their highly coupled nature. At the most resonant of the computed combinations ($g=0.999$ and $a_1=0.001$), the RT60 is 86.26s at $f_0=80$Hz, while being only 4.5s at $f_0=1280$Hz.}
    \label{fig:rt60}
\end{figure}

For RT60 ($\mathcal{L}_{\text{rt60}}$), rather than supervising the coupled $g$ and $a_1$ independently, we compute their joint analytical decay time. A signal at harmonic $k$ circulates through $H_L(e^{j\omega_k})$ exactly $f_0$ times per second, giving:
\begin{equation}
\label{eq:rt60}
\text{RT60}_k = \frac{-3}{f_0 \log_{10}\lvert H_L(e^{j\omega_k})\rvert},
\end{equation}
where the $-3$ in the numerator reflects the $-60$\,dB (factor $10^{-3}$) amplitude decay that defines RT60, evaluated for the first 10 harmonics below Nyquist. An $L_1$ loss in $\log(1+\text{RT60})$ space provides balanced gradients across short and long decays.

\subsection{Data}\label{sec:data}

\textbf{Synthetic Dataset (Synth)}. Samples are 4-second mono clips at $f_s{=}16$\,kHz containing 1--10 pluck events at constant \gls{f_0} over D1--D6, with per-event parameters sampled from Beta distributions and a 30\% palm-mute probability. Full generation details and reproducible code can be found in appendix \ref{sec:appendix}. The training set comprises 8360 samples re-seeded each epoch; a fixed 290-sample validation set monitors convergence.

\textbf{Real-World Dataset (Real)}. We use the acoustic guitar subset of NSynth~\cite{Engel:2017:Neural}, retaining non-reverberant samples in the D1--D6 range, yielding 4-second mono recordings at 16\,kHz split as [8360\,/\,930\,/\,290] for train\,/\,validation\,/\,test.

\subsection{Models}\label{sec:baselines}
We evaluate seven models across three regimes. \textbf{$\mathcal{P}$-Only} uses the non-differentiable tKSA trained with \gls{p_loss} alone, predicting all event parameters (existence, onset, $f_0$, timbre) end-to-end. \textbf{Audio-Only} models use differentiable \gls{ksa} decoders with audio losses only, replacing the existence, onset, and $f_0$ heads with external detectors: CREPE~\cite{Kim:2018:Crepe} (Viterbi, tiny) for $f_0$ and Spectral Flux with backtracking~\cite{Mcfee:2015:Librosa} for onsets; the signal is segmented at detected onsets and each event's $f_0$ set to the median over its confidently voiced frames (CREPE confidence $>0.5$). \textbf{$\mathcal{P}$+Audio} models use the same differentiable decoders trained end-to-end with both \gls{p_loss} and audio losses (\gls{mss}, \gls{sot}), jointly predicting all parameters; we further ablate them by detaching the audio-loss gradient from the onset, $f_0$, or both heads. All \gls{ksa} variants are evaluated in both frequency-sampling (freq.) and time-domain (time) forms. \textbf{HpN} and \textbf{HpN+} are \acrlong{hn} baselines using the original \gls{ddsp} encoder~\cite{Engel:2020:DDSP} and our \gls{tcn} encoder respectively, both trained with audio losses and external detectors (CREPE for $f_0$, A-weighted loudness for dynamics), without reverberation.

\subsection{Training Process}\label{sec:training-process}
All models use AdamW ($\text{lr}{=}10^{-3}$, weight decay $10^{-4}$, batch size 32) with gradient clipping at norm 1.0 and ReduceLROnPlateau (factor 0.5, patience 15, min $\text{lr}{=}10^{-6}$), on A100 GPUs\footnote{This research utilised Queen Mary's Apocrita HPC facility, supported by QMUL Research-IT. http://doi.org/10.5281/zenodo.438045} with early stopping (patience 50, threshold $10^{-4}$, max 24\,hrs or 10000 epochs). $\mathcal{P}$+Audio training adds the parameter loss at $w_{\mathcal{P}}{=}1.0$ to the audio-loss weights of §\ref{sec:losses} ($w_{\text{MSS}}{=}0.05$, $w_{\text{SOT}}{=}1.0$), which Audio-Only models use alone. When mixing datasets, epoch sizes are matched and dataloaders combined via max-size cycling.

\section{Evaluation}
\label{sec:evaluation}
We evaluate on two 290-sample test sets: held-out synthetic and NSynth acoustic guitar. For Audio-Only models on synthetic data, detectors run on-the-fly to mirror training. Metrics fall into three categories.

\textbf{Signal Fidelity}. To assess signal fidelity, we employ \acrshort{mss} and \acrshort{sot} identically to their usage during training (Eqs.~\ref{eq:mss},~\ref{eq:total-loss}). We also evaluate perceptual timbral distance using a warped \gls{mfcc} metric, which calculates the \gls{dtw}-normalised $L_1$ distance between 20-coefficient \gls{mfcc} sequences; the dynamic time warping naturally absorbs minor temporal misalignments. Additionally, we measure the cosine similarity between the target and predicted \gls{rms} energy envelopes to evaluate the overall envelope fidelity.

\textbf{Parameter and Event Reconstruction}. Because ground-truth parameters are required, these metrics are computed exclusively on the synthetic test set. We evaluate event detection performance using onset precision, recall, and F1 scores by matching predicted slots that have an existence probability greater than $0.5$ against ground-truth onsets within a 25ms window. Pitch accuracy is measured by calculating the MAE $f_0$ in cents ($\Delta$ Cents) across all matched active events. Finally, we report the overall parameter loss ($\mathcal{P}_{\text{Loss}}$) as previously defined in Eq.~\ref{eq:param-loss}.

\textbf{Distributional and Perceptual Similarity}. For the real-world NSynth dataset, we compute \gls{kad}~\cite{Chung:2025:Kad}, a distribution-free Maximum Mean Discrepancy metric that, unlike the Fréchet Audio Distance, remains strictly unbiased even at small sample sizes. Following the methodology of \cite{Tailleur:2024:Correlation}, we calculate KAD across two distinct embedding spaces to capture different perceptual qualities: EnCodec~\cite{Defossez:2023:High} is used to measure low-level spectro-temporal fidelity, while CLAP~\cite{Elizalde:2023:Clap} provides an assessment of higher-level semantic and timbral alignment.

\section{Experiments}
\label{sec:results}
\textbf{Gradient Analysis.} We evaluate whether gradients point toward the target parameter:\begin{equation}\text{sgn} \left( \frac{\partial \mathcal{L}(\hat{x}f[n], x[n])}{\partial f} \right) = \text{sgn}(f - f_x)
\end{equation}
Table~\ref{tab:single} reports Coarse Gradient Accuracy (CGA, 250 uniform initialisations across the full parameter range) and Fine Gradient Accuracy (FGA, 250 initialisations within $\pm10\%$ of target) for a single event at $\{f_0, t, a_1, g, p, \text{gain}, \text{dyn}\} = \{110\,\text{Hz}, 2\,\text{s}, 0.2, 0.99, 0.25, 0.9, 0.9\}$, across both losses ($\mathcal{L}_{\text{MSS}}$, $\mathcal{L}_{\text{SOT}}$) and implementations (tKSA, fKSA). Table~\ref{tab:multi} extends this to multi-event sequences ($K \in [2,5]$) focusing on the onset time parameter. To maintain a consistent evaluation budget of approximately 250 initialisations regardless of sequence length, we construct a uniform $K$-dimensional grid where the number of test points per dimension scales as $\lfloor 250^{1/K} \rceil$ (e.g., $16^2=256$ for $K=2$, $6^3=216$ for $K=3$). We report \emph{Any} accuracy (gradient points toward any ground-truth event) and \emph{Joint} accuracy (all $K$ predictions simultaneously pulled toward their Hungarian-matched targets).

\textbf{Synthetic Parameter Recovery.} All five \gls{ksa} models (§\ref{sec:baselines}) are trained and evaluated on synthetic data, scored with signal-fidelity and parameter-reconstruction metrics (§\ref{sec:evaluation}); results are reported in Table~\ref{tab:synth}.

\textbf{Real-World Instrument Modelling.} $\mathcal{P}$+Audio models train on mixed Synth+Real data, Audio-Only and the \gls{hn} baselines on Real data; $\mathcal{P}$-Only is unchanged from the previous experiment. Evaluation uses the NSynth acoustic guitar test set with signal-fidelity and distributional metrics (§\ref{sec:evaluation}); results are reported in Table~\ref{tab:real}.

\begin{table}[t!]
\centering
\caption{Single-event gradient accuracy (\%) comparing $\mathcal{L}_{MSS}$ and $\mathcal{L}_{SOT}$ across tKSA and fKSA.}
\label{tab:single}
\resizebox{\columnwidth}{!}{%
\begin{tabular}{@{}lllllllll@{}}
\toprule
          & \multicolumn{4}{c}{$\mathcal{L}_{MSS}$} & \multicolumn{4}{c}{$\mathcal{L}_{SOT}$} \\ 
          & \multicolumn{2}{c}{tKSA}       & \multicolumn{2}{c}{fKSA}      & \multicolumn{2}{c}{tKSA}       & \multicolumn{2}{c}{fKSA}      \\
Parameter & CGA     & FGA     & CGA     & FGA    & CGA    & FGA     & CGA    & FGA    \\ \midrule
$f_0$           & 68.0  & 58.4  & 69.6  & 63.6 & 54.8 & 50.8  & 50.0 & 50.0 \\
$g$ (decay)     & 100.0 & 100.0 & 100.0 & 99.6 & 70.4 & 47.6  & 73.6 & 47.6 \\
$a_1$ (damping) & 100.0 & 100.0 & 99.2  & 94.8 & 40.0 & 50.0  & 34.4 & 50.0 \\
Pluck position  & 66.8  & 89.2  & 67.6  & 90.0 & 49.6 & 48.0  & 47.6 & 46.0 \\
Dynamic level   & 100.0 & 99.6  & 98.4  & 92.4 & 97.6 & 100.0 & 98.0 & 98.8 \\
Distance gain      & 100.0 & 100.0 & 96.4  & 82.0 & 99.6 & 99.2  & 97.2 & 86.8 \\
Onset time       & 51.2  & 55.2  & 52.0  & 56.0 & 42.4 & 14.4  & 62.4 & 88.8 \\ \bottomrule
\end{tabular}%
}
\end{table}
\begin{table}[t!]
\centering
\caption{Multi-event onset time gradient accuracy (\%) for $K$ equally spaced plucks. fKSA uses $N_{FFT}=16384$, hop=256. \emph{Any}: at least one gradient points toward a target; \emph{Joint}: all $K$ simultaneously toward their Hungarian-matched targets.}

\label{tab:multi}
\resizebox{\columnwidth}{!}{%
\begin{tabular}{@{}l cc cc cc cc@{}}
\toprule
 & \multicolumn{4}{c}{$\mathcal{L}_{MSS}$} & \multicolumn{4}{c}{$\mathcal{L}_{SOT}$} \\ 
 & \multicolumn{2}{c}{tKSA} & \multicolumn{2}{c}{fKSA} & \multicolumn{2}{c}{tKSA} & \multicolumn{2}{c}{fKSA} \\ 
Events & Any & Joint & Any & Joint & Any & Joint & Any & Joint \\ \midrule
2 & 73.4 & 22.7 & 77.1 & 23.0 & 64.1 & 12.1 & 77.5 & 28.5 \\
3 & 83.3 & 7.4  & 82.7 & 13.9 & 79.9 & 5.6  & 84.9 & 11.6 \\
4 & 71.6 & 1.6  & 72.8 & 4.7  & 70.7 & 1.2  & 77.6 & 6.6  \\
5 & 65.9 & 0.4  & 65.3 & 3.3  & 66.9 & 1.2  & 68.6 & 4.1  \\ \bottomrule
\end{tabular}%
}
\end{table}
\begin{figure}[t!]
    \centering
    \includegraphics[width=\columnwidth]{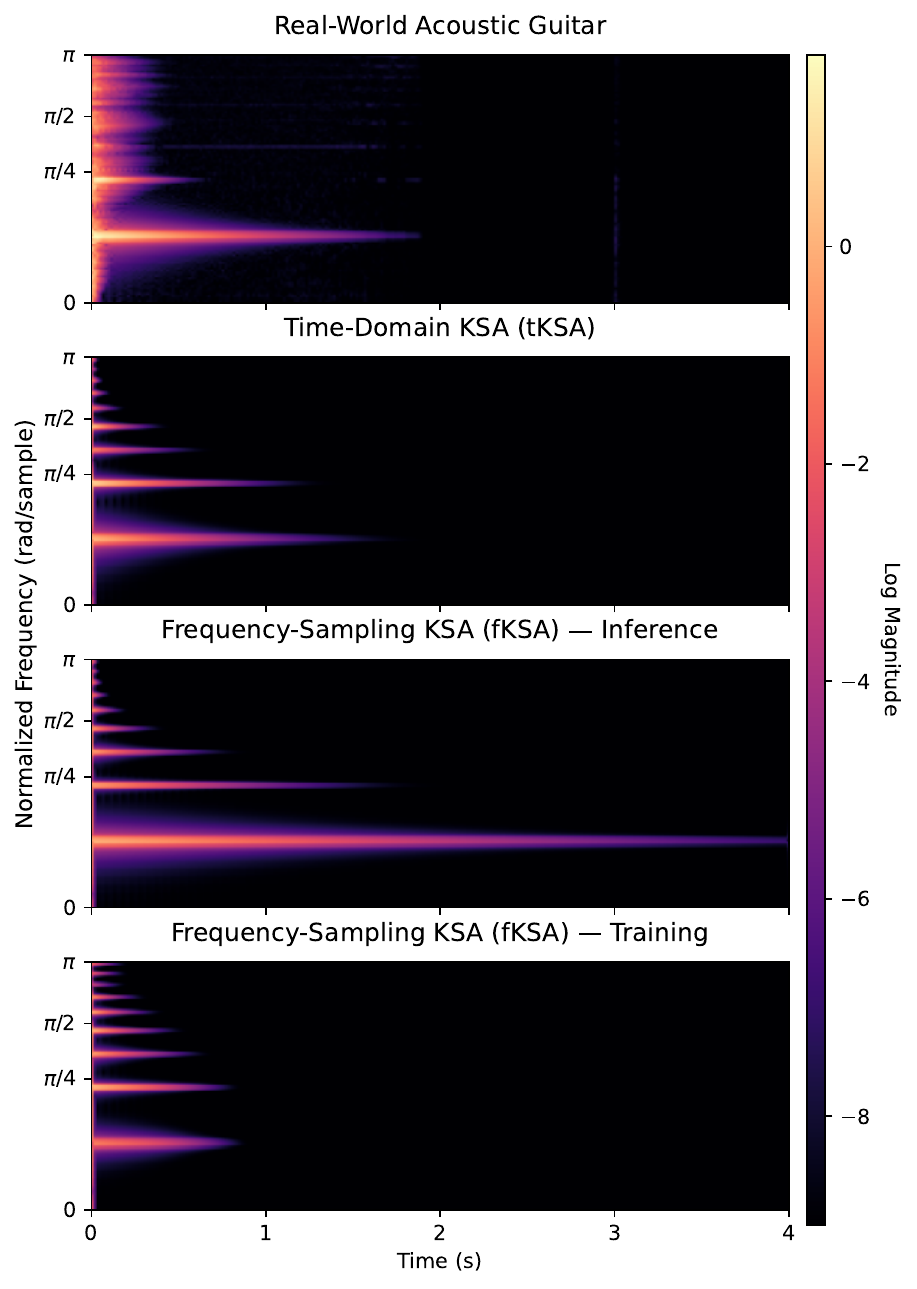}
\caption{Sound matching on an NSynth acoustic guitar recording. Top: target.
Middle: tKSA and fKSA Audio-Only reconstructions at inference. Bottom: the same
fKSA reconstruction rendered via the frequency-sampling forward pass used in
training, where time-aliasing from circular convolution at $N_{FFT}=2^{14}$
truncates long decays, leading the model to overestimate decay $g$ and
underestimate damping $a_1$, visible as excess sustain above.}
    \label{fig:nsynth_tKSA_vs_fKSA}
\end{figure}
\begin{table*}[h!]
\centering
\caption{Synthetic parameter recovery. All models are trained and evaluated on synthetic data. $\mathcal{P}$-Only uses tKSA with $\mathcal{P}_{\text{Loss}}$ supervision; $\mathcal{P}$+Audio models use differentiable decoders trained end-to-end with both $\mathcal{P}_{\text{Loss}}$ and audio losses ($\mathcal{L}_{\text{MSS}}$, $\mathcal{L}_{\text{SOT}}$); Audio-Only models use differentiable decoders with audio losses and external detectors for onsets and $f_0$. Rows beneath each $\mathcal{P}$+Audio model are stop-gradient ablations that detach the audio-loss gradient from the named head ($f_0$, onset, or both), leaving it supervised only by $\mathcal{P}_{\text{Loss}}$. Per column, the best, second-best, and third-best values are shown in \textbf{bold}, \underline{underline}, and \textit{italic} respectively.}
\label{tab:synth}
\begin{tabular}{lrrrrrrrrr}
\toprule
 & \gls{mss} $\downarrow$ & \gls{sot} $\downarrow$ & wMFCC $\downarrow$ & RMS $\uparrow$ & Precision $\uparrow$ & Recall $\uparrow$ & F1 $\uparrow$ & $\Delta$ Cents $\downarrow$ & \gls{p_loss} $\downarrow$ \\
\midrule
$\mathcal{P}$-Only & \textbf{136.11} & \underline{0.18} & \textbf{6.02} & \textbf{0.96} & \underline{0.92} & \textbf{0.92} & \textbf{0.92} & 15.14 & \textbf{0.29} \\
\midrule
Audio-Only (freq.) & 158.95 & \textit{0.19} & 9.29 & 0.91 & \textbf{0.94} & \underline{0.85} & \underline{0.89} & 294.39 & 17.49 \\
Audio-Only (time) & \underline{149.07} & \textbf{0.16} & 9.23 & \textit{0.93} & \textbf{0.94} & \underline{0.85} & \underline{0.89} & 294.33 & 15.66 \\
\midrule
$\mathcal{P}$+Audio (freq.) & 252.90 & \textit{0.19} & 15.29 & 0.33 & 0.01 & 0.01 & 0.01 & 116.96 & 4.72 \\
\quad$\hookrightarrow$ detach $f_0$ & 251.79 & 0.22 & 15.09 & 0.36 & 0.02 & 0.02 & 0.02 & 45.49 & 3.73 \\
\quad$\hookrightarrow$ detach onset & 195.29 & 0.27 & 10.92 & 0.81 & 0.31 & 0.31 & 0.31 & 42.60 & 1.94 \\
\quad$\hookrightarrow$ detach both & 163.00 & 0.21 & \textit{8.15} & \underline{0.94} & 0.83 & 0.83 & 0.83 & \textbf{12.69} & \textit{0.52} \\
$\mathcal{P}$+Audio (time) & 241.16 & 0.20 & 13.47 & 0.43 & 0.01 & 0.01 & 0.01 & \textit{13.86} & 3.74 \\
\quad$\hookrightarrow$ detach $f_0$ & 245.77 & \underline{0.18} & 16.29 & 0.38 & 0.01 & 0.01 & 0.01 & 216.76 & 4.55 \\
\quad$\hookrightarrow$ detach onset & 157.95 & 0.20 & 8.19 & \textit{0.93} & 0.64 & 0.64 & 0.64 & \underline{13.15} & 0.82 \\
\quad$\hookrightarrow$ detach both & \textit{155.96} & 0.23 & \underline{7.59} & \underline{0.94} & \textit{0.84} & \textit{0.84} & \textit{0.84} & 14.26 & \underline{0.48} \\
\bottomrule
\end{tabular}
\end{table*}
\begin{table}[t]
\centering
\caption{Real-world instrument modelling on the NSynth acoustic-guitar test set. \gls{ksa} regimes are as in Table~\ref{tab:synth}, but with $\mathcal{P}$+Audio trained on mixed Synth+Real data and Audio-Only on Real data; HpN and HpN+ are \acrlong{hn} baselines (original \gls{ddsp} and our \gls{tcn} encoder) trained on Real data. KAD is reported in CLAP and EnCodec embedding spaces. Ablations and highlighting follow Table~\ref{tab:synth}.}
\label{tab:real}
\resizebox{\ifdim\width>\columnwidth\columnwidth\else\width\fi}{!}{%
\begin{tabular}{lrrrrrr}
\toprule
 & \rotatebox[origin=c]{-90}{\gls{mss}}\,$\downarrow$ & \rotatebox[origin=c]{-90}{\gls{sot}}\,$\downarrow$ & \rotatebox[origin=c]{-90}{wMFCC}\,$\downarrow$ & \rotatebox[origin=c]{-90}{RMS}\,$\uparrow$ & \rotatebox[origin=c]{-90}{CLAP}\,$\downarrow$ & \rotatebox[origin=c]{-90}{EnCodec}\,$\downarrow$ \\
\midrule
HpN & \underline{246.60} & \textit{0.14} & \underline{7.90} & \textit{0.96} & 266.20 & \underline{17.99} \\
HpN+ & \textbf{183.43} & \textbf{0.02} & \textbf{6.20} & \textbf{1.00} & \underline{183.11} & \textbf{5.20} \\
\midrule
$\mathcal{P}$-Only & 311.86 & 0.20 & 13.18 & 0.95 & 193.78 & 64.40 \\
\midrule
Audio-Only (freq.) & 299.47 & 0.30 & 14.08 & 0.93 & \textit{191.50} & 71.91 \\
Audio-Only (time) & \textit{262.72} & 0.20 & \textit{12.18} & \underline{0.97} & \textbf{163.95} & \textit{40.39} \\
\midrule
$\mathcal{P}$+Audio (freq.) & 468.33 & \underline{0.13} & 32.43 & 0.09 & 547.78 & 494.88 \\
\quad$\hookrightarrow$ detach $f_0$ & 466.40 & 0.20 & 30.80 & 0.12 & 431.52 & 482.48 \\
\quad$\hookrightarrow$ detach onset & 393.13 & 0.28 & 21.44 & 0.86 & 363.81 & 118.33 \\
\quad$\hookrightarrow$ detach both & 348.18 & 0.25 & 15.37 & 0.81 & 204.19 & 116.66 \\
$\mathcal{P}$+Audio (time) & 461.06 & 0.18 & 26.63 & 0.16 & 403.83 & 470.67 \\
\quad$\hookrightarrow$ detach $f_0$ & 463.00 & 0.18 & 27.29 & 0.16 & 446.45 & 480.10 \\
\quad$\hookrightarrow$ detach onset & 336.72 & 0.19 & 14.88 & 0.82 & 213.53 & 96.80 \\
\quad$\hookrightarrow$ detach both & 341.44 & 0.21 & 15.08 & 0.83 & 201.14 & 99.01 \\
\bottomrule
\end{tabular}%
}
\end{table}

\section{Discussion}
\label{sec:discussion}
Table~\ref{tab:single} confirms that time-domain interpolation gradients are comparable to frequency-sampling, validating tKSA as a viable alternative. As a \emph{vertical}, point-wise distance, \gls{mss} provides reliable gradients for timbral and dynamics parameters but struggles with pluck position. The \emph{horizontal} \gls{sot}, which only redistributes energy along the frequency axis, is weaker on most parameters and, surprisingly given its design, blind to $f_0$; we hypothesise this is due to the fact that \gls{ksa} delay length influences the decay, as shown in Figure~\ref{fig:rt60}. Its one striking exception is onset time under fKSA, where it reaches the highest single-event accuracy of any configuration (88.8\% FGA), against near-blindness on the alias-free tKSA (14.4\%). A plausible explanation is that frequency-sampling time-aliasing (Fig.~\ref{fig:nsynth_tKSA_vs_fKSA}) couples a temporal shift into the per-frame spectral content that the horizontal \gls{sot} can resolve, an artefact rather than a robust onset cue; however further experiments would be needed to confirm this.

Critically, this single-event advantage does not survive to realistic sequences. Table~\ref{tab:multi} reports \emph{Joint} onset accuracy, all $K$ onsets pulled toward their matched targets at once, collapsing for every loss as $K$ grows (to at most $4.1\%$ at $K{=}5$), as the losses distribute events unevenly and yield unusable temporal structure. No audio loss, then, provides usable gradients for the onsets of a multi-event signal. This explains the consistent underperformance of $\mathcal{P}+$Audio ablations relative to $\mathcal{P}$-Only and Audio-Only counterparts across both experiments, evidenced by the low RMS values\footnote{We hypothesise the low \gls{sot} in these collapsed models is an artefact of per-frame low-pass filtering normalisation, resulting in unreliable metrics for reconstructions with wrongly-predicted silent frames.}. This is corroborated by the $\mathcal{P}+$Audio ablations, where detaching the onset head from audio losses always results in strongly increased performance. We additionally hypothesise that time-aliasing is the primary factor further degrading frequency-sampling relative to time-domain implementations in Audio-Only variants, where identical onset and $f_0$ targets isolate the aliasing effect. As shown in Fig.~\ref{fig:nsynth_tKSA_vs_fKSA}, decays whose impulse responses exceed the FFT frame length are truncated by circular convolution, causing the fKSA model to underestimate sustain.

The $\mathcal{P}$-Only ablation performs best on synthetic data, even for timbre. In addition to being given a clear direction towards the global optima, we attribute this to the perceptual RT60 reparameterisation (see Eq. \ref{eq:rt60}), which jointly supervises damping and decay. $\Delta$ Cents error is inflated by octave misalignments in Audio-Only models, which also show lower onset recall. 

However, \gls{p_loss} generalises poorly to real-world data, frequently missing onsets, likely due to CNN overfitting to synthetic transient features and erratically predicting $f_0$. Future work could narrow this gap by augmenting the synthetic data with a realistic noise floor and phase distortion from randomised all-pass filters to emulate microphone capture, by supervising $f_0$ and onset times with a higher-fidelity physical simulation such as a finite-difference time-domain model, or, most directly, by training on annotated real-world recordings.

As expected, the \gls{hn} baselines achieve superior reconstruction across most metrics. Unsurprisingly, this gap reflects a structural ceiling of the \gls{ksa}: its single feedback loop captures neither the resonant guitar body, nor the interaction between the string and the player's hands, nor string--fret collisions, all contributions the \gls{hn} model can absorb directly and that future research should address through more comprehensive physical models. Although we argue that our proposed \gls{ksa} decoder represents transient characteristics more faithfully than the \gls{hn} variants, we acknowledge that increasing the frame resolution of \gls{hn} is likely to further improve its ability to reconstruct transients and that no formal listening test was conducted. Instead, we encourage readers to listen to the audio examples in the accompanying repository (appendix \ref{sec:appendix}). 

Lastly, designing audio losses that yield useful gradients for event \emph{location}, and for $f_0$, remains a key open challenge. Vertical, point-wise distances cannot indicate which way spectral energy should move; optimal-transport formulations, in theory, should. Beyond \gls{sot} along the frequency axis~\cite{Torres:2024:Unsupervised}, a \emph{joint} time--frequency transport cost~\cite{Fabiani:2024:Time}, growing with displacement in both time and frequency, could in principle resolve onset and $f_0$ \emph{together}.

\section{Conclusion}
We presented an event-based sound-matching system built on a differentiable extended Karplus-Strong algorithm, comparing time-domain and frequency-sampling decoders under end-to-end and external-detector training. Time-domain Lagrange interpolation matches frequency-sampling gradient accuracy while avoiding frame-based time-aliasing, making it a viable, artefact-free alternative for differentiable physical modelling of time-varying highly-resonant systems.

Our gradient analysis identifies the absence of directional onset gradients from spectral losses as the primary failure mode of joint audio-parameter training. Parameter loss on synthetic data supervises $f_0$, timbre, and onsets well, but learned detectors do not yet generalise to real transients; external detectors with audio losses remain most robust on real data, though Harmonics-plus-Noise baselines retain higher reconstruction fidelity on most metrics. The exception is the CLAP perceptual distance, on which our time-domain Audio-Only model surpasses both baselines; we hypothesise this stems from the \gls{ksa}'s natural modelling of sharp transients, though formal listening tests are needed to confirm it. All audio examples are in appendix \ref{sec:appendix}.

Future work should close the sim-to-real onset gap via training on higher-quality synthetic and annotated acoustic data, and explore joint time-frequency optimal-transport losses, which measure spectral energy displacement along both axes simultaneously \cite{Fabiani:2024:Time, Torres:2024:Unsupervised}. Extending the decoder to a more comprehensive physical model that simulates string-fret collisions and string-player interactions is the natural path toward higher-fidelity guitar reconstruction.

\appendix
\section{Audio Examples and Code}
\label{sec:appendix}
Audio examples for all synthetic parameter recovery and real-world instrument modelling experiments, along with the reference code for reproducing experiments, figures, synthetic data generation, and NSynth data filtering, are available at \url{https://ptablasdpaula.github.io/DAFx26-Karplus/}.

\bibliographystyle{IEEEtranDAFx}
\bibliography{DAFx26_tmpl} 

\end{document}